\documentclass[twocolumn, twocolappendix, apj, apjl]{theoj}
\usepackage{amsmath}
\usepackage{booktabs}
\usepackage{bm}
\usepackage{physics}
\usepackage{hyperref}

\shorttitle{Score-Based Reconstruction of Galaxy Cluster Mass Maps}
\shortauthors{Hsu et al.}

\begin{document}

\title{Reconstructing Galaxy Cluster Mass Maps using Score-based Generative Modeling}

\author{
Alan Hsu\(^{1,2}\),
Matthew Ho\(^{3,4}\),
Joyce Lin\(^{5,2}\),
Carleen Markey\(^{2,6}\),
Michelle Ntampaka\(^{7,8}\),
Hy Trac\(^{2,6}\),
Barnab\'{a}s P\'{o}czos\(^{9}\)
}

\affiliation{\(^1\)Department of Astronomy, Harvard University, Cambridge, MA 02138, USA}
\affiliation{\(^2\)Department of Physics, Carnegie Mellon University, Pittsburgh, PA 15213, USA}
\affiliation{\(^3\)CNRS \& Sorbonne Universit\'{e}, Institut d’Astrophysique de Paris (IAP),
UMR 7095, 98 bis bd Arago, F-75014 Paris, France}
\affiliation{\(^4\)Columbia Astrophysics Laboratory, Columbia University, 550 West 120th Street, New York, NY 10027, USA}
\affiliation{\(^5\)Department of Physics, University of Wisconsin-Madison, Madison, Wisconsin 53726, USA}
\affiliation{\(^6\)McWilliams Center for Cosmology and Astrophysics, Carnegie Mellon University, Pittsburgh, PA 15213, USA}
\affiliation{\(^7\)Data Science Mission Office, Space Telescope Science Institute, Baltimore, MD, 21218, USA}
\affiliation{\(^8\)Department of Physics \& Astronomy, Johns Hopkins University, Baltimore, MD 21218, USA}
\affiliation{\(^9\)Machine Learning Department, Carnegie Mellon University, Pittsburgh, PA 15213, USA}

\begin{abstract}

We present a novel approach to reconstruct gas and dark matter projected density maps of galaxy clusters using score-based generative modeling. Our diffusion model takes in mock SZ and X-ray images as conditional inputs, and generates realizations of corresponding gas and dark matter maps by sampling from a learned data posterior. We train and validate the performance of our model by using mock data from a cosmological simulation. The model accurately reconstructs both the mean and spread of the radial density profiles in the spatial domain, indicating that the model is able to distinguish between clusters of different mass sizes. In the spectral domain, the model achieves close-to-unity values for the bias and cross-correlation coefficients, indicating that the model can accurately probe cluster structures on both large and small scales. Our experiments demonstrate the ability of score models to learn a strong, nonlinear, and unbiased mapping between input observables and fundamental density distributions of galaxy clusters. These diffusion models can be further fine-tuned and generalized to not only take in additional observables as inputs, but also real observations and predict unknown density distributions of galaxy clusters.

\end{abstract}

\section{Introduction} \label{sec:intro}

Galaxy clusters are the largest gravitationally bound systems in the universe, typically comprised of hundreds of galaxies and a hot intracluster medium (ICM) embedded in a dark matter halo. The cluster mass scales with galaxy velocity dispersion, a relation famously used to postulate the presence of dark matter in the Coma cluster \citep[][]{zwicky}. The number density of galaxy clusters as a function of their mass is a cosmological probe, and is sensitive to the distribution and evolution of large-scale structures, as density is a probe into the growth of structure which depends on cosmological parameters (see \citet{2011ARA&A..49..409A} and \citet{2012ARA&A..50..353K} for reviews).

The primary baryonic component of galaxy clusters is the ICM, a hot, ionized gas that is observable over a range of wavelengths. The plasma radiates through thermal bremsstrahlung (free-free) emission at X-ray energies. The free electrons upscatter the cosmic microwave background (CMB) through the Sunyaev-Zel'dovich \citep[SZ,][]{1972CoASP...4..173S} effect and is detected at microwave frequencies. Observations of the ICM can be used as a mass proxy, a way to infer the underlying mass from observables, and both X-ray \citep[e.g.][]{2006ApJ...640..691V, Pratt_2009, Mantz_2016, Giles_2017} and SZ \citep[e.g.][]{2011ApJ...737...61M, 2013ApJ...763..127R, 2021ApJS..253....3H, 2021A&A...647A.106M, 2024OJAp....7E..13B} observations of clusters have been utilized in this way.  These mass proxies may either be derived from first principles or calibrated on cluster simulations \citep[e.g.][]{Kravtsov_2006, Kay_2012, 2023MNRAS.522.2628W}.

Recently, machine learning models have been shown to improve standard estimates of galaxy cluster mass, using observables such as line-of-sight velocities and projected radial distances \citep[e.g.][]{2019ApJ...887...25H, 2020MNRAS.499.1985K, 2021MNRAS.501.4080K, 2021ApJ...908..204H, 2022NatAs...6..936H}, X-ray \citep[e.g.][]{Ntampaka_2019, Green_2019, Han_2020, Ho_2023} and SZ \citep[e.g.][]{Cohn_2019, Wadekar_2023_1}. However, compared to standard estimates of cluster mass, estimating the projected 2D mass map is even more informative: knowing the mass map at different radii not only estimates cluster abundance but also better constrains lensing maps and dark matter models. Finally, since clusters effectively act as giant telescopes to magnify and find high redshift galaxies and quasars, having a model that accurately produces the mass map is crucial. To do so, we will need a machine learning model that can produce images rather than single-number estimates for the mass. 

Concurrently, image-to-image deep learning models, notably autoencoder-style networks such as Generative Adversarial Models (GANs) and Variational Autoencoders (VAEs), are very good at solving ill-posed astrophysics problems that require learning a compressed latent space representation and sampling from a learned distribution: these models learn to capture and appropriately model the intrinsic uncertainty in incomplete observations. Examples of reconstruction from the latent space include random sampling of hyper-realistic large cosmological structures \citep[e.g.][]{Ullmo_2021}, and accurate reconstruction of galaxy images \citep[e.g.][]{Schawinski_2018} and large-scale structures \citep[e.g.][]{Rodriguez_2018, Perraudin_2019}. Other interesting applications of image generation include reconstructing total projected matter maps from cluster observables \cite[e.g.][]{Andres_2024}, classification of galaxy mergers \cite[e.g.][]{Arendt_2024}, denoising and deconvolving galaxy images \citep[e.g.][]{Schawinski_2017, Hemmati_2022} and increasing the resolution of cosmological simulations \citep[e.g.][]{Li_2021}.

A recent paradigm shift in image-to-image generative modeling is the development of diffusion or score-based generative models (SGMs), which is a probabilistic framework that reformulates the previous sampling processes from learned data distributions. By increasingly injecting noise into our data and learning the inverse denoising procedure, we can map a prior distribution to the data distribution and subsequently sample from it. Score matching with Langevin Dynamics \citep[SMLD, e.g.][]{Vincent_2011, Song_2019, Song_2021} learns the score, or the gradient of the log probability density, of the data distribution which is used by Langevin dynamics to generate data using stochastic differential equations (SDEs). An analogous score-based framework is Denoising Diffusion Probabilistic Modeling \citep[DDPM, e.g.][]{Sohl-Dickstein_2015, Ho_2020} which implicitly learns the score by training a sequence of models to reverse the noise injection. SGMs offer many advantages over traditional GANs and VAEs: not only are SGMs less prone to mode collapse to an average solution during training, they also learn the noise of the problem at all scales in a structured manner, which is crucial for astronomical data that may be noisy, incomplete, or lossy. SGMs also optimize using an explicit likelihood surrogate, whose stability scales well with dimension, as opposed to the amortized inference procedure of VAEs that can under-represent the posterior, or the adversarial losses used by GANs that is known to be unstable. In the context of astrophysics generative modeling, diffusion models have been able to robustly sample various posterior distributions: recent work in astrophysics has been applied to a variety of tasks, including sampling galaxy spectra \citep[e.g.][]{Doorenbos_2022, Doorenbos_2024}, super-resolution gravitational lensing maps \citep[e.g.][]{Reddy_2024}, satellite galaxies and subhalo populations \citep[e.g.][]{Nguyen_2024, bourdin2024inpainting}, and initial conditions of the universe \citep[e.g.][]{Legin_2024}.

We present a score-based generative model that learns the predictive posterior distribution of gas and dark matter maps of galaxy clusters. This powerful probabilistic model allows us to subsequently sample realizations of these maps conditioned on SZ and X-ray inputs. In Section \ref{sec:mock_generation}, we detail the process of generating mock observations from a cosmological simulation as training and testing data for our models. In section \ref{sec:machine_learning}, we describe the diffusion model pipeline and elaborate on the statistical framework of learning the score of the data distribution. Finally, in section \ref{sec:results} we present the estimated gas and dark matter maps, the reconstructed accuracy of the density profiles in the spatial domain, and the bias and cross correlation coefficients in the spectral domain. 
\section{Simulations} \label{sec:mock_generation}

We apply score-based generative modeling on simulated galaxy clusters to study its potential for reconstructing the underlying mass maps. With simulated galaxy clusters, we know the gas and dark matter properties and distributions and can construct corresponding multi-wavelength mock observations. This important, pilot step to quantify the accuracy and uncertainties with simulated images is necessary before the approach can be applied to real observations. In this section, we first describe the cosmological simulation (Section \ref{sec:sim}) and galaxy cluster sample (Section \ref{sec:clusters}). We then describe the construction of mock observations of the SZ effect (Section \ref{sec:sz}) and X-ray emission (Section \ref{sec:xray}).

%%%%%%%%%%%%%%%%%%%%%%%%%%%%%%%%%%%%%%%%%
\subsection{Cosmological Simulation} 
\label{sec:sim}
%%%%%%%%%%%%%%%%%%%%%%%%%%%%%%%%%%%%%%%%%

Many supervised machine learning models, such as the image-to-image diffusion model that will be described in Section \ref{sec:machine_learning}, require large amounts of data for training. We generate a large number of simulated galaxy clusters by running a cosmological simulation with an updated version of \texttt{HYPER} \citep{2022ApJ...925..134H}, a hydro-particle code for efficient and rapid simulations of baryons and dark matter. The hydro simulation used here has a moderately large volume so as to produce many galaxy clusters, paired with a high enough resolution to model internal cluster structures. This combination of large volume and high resolution is made possible due to the novel use of subgrid models for the thermodynamics of the ICM and intergalactic medium (IGM), resulting in a three-order-of-magnitude speedup compared to a standard hydrodynamics code.

We summarize the recent updates to $\texttt{HYPER}$, which will be presented in more detail in an upcoming paper. For the gravity solver, the particle-mesh (PM) algorithm is upgraded to particle-particle-mesh \citep[P$^3$M;][]{2015ApJ...813...54T} for improved spatial resolution. For the hydro solver, some components of the hydro-particle-mesh \citep[HPM;][]{1998MNRAS.296...44G, 2022ApJ...925..134H} algorithm are replaced with elements from smoothed particle hydrodynamics \citep[SPH; see][for a review]{2010ARA&A..48..391S}. Particle densities and pressure gradients are not calculated using a mesh, but rather by summing over and weighting particles within the individual smoothing lengths. The kick-drift-kick leapfrog integrator now uses adaptive time steps for improved efficiency.

To generate mock SZ and X-ray observations of galaxy clusters, we simulate the ICM using subgrid models. These models include a dark matter halo model, an ICM pressure profile, and an IGM temperature-density relation, all of which can be systematically varied for parameter space studies. In particular, the ICM pressure profile is based on the debiased pressure profile \citep[DPP;][]{2021ApJ...908...91H}, which accounts for hydrostatic mass bias by combining results from X-ray observations \citep[e.g.][]{2007A&A...469..363B} with cosmological simulations \citep[e.g.][]{2021MNRAS.506.2533B}. In \citet{2022ApJ...925..134H}, the $\texttt{HYPER}$ simulation results agree well with the halo model expectations for the density, temperature, and pressure radial profiles. Additional details are provided in Appendix \ref{sec:haloprofiles}.

The simulation is based on the following cosmological parameters: $\Omega_\mathrm{b} = 0.045$, $\Omega_\mathrm{m} = 0.3$, $\Omega_\Lambda = 0.7$, $h=0.7$, $\sigma_8 = 0.8$, and $n_\mathrm{s} = 0.96$. There are $N_\mathrm{dm}=1024^3$ dark matter particles and $N_\mathrm{gas}=1024^3$ gas particles in a comoving box of side length $L_\mathrm{sim} = 500\ h^{-1}\mathrm{Mpc}$. The particle masses are $m_\mathrm{dm} = 1.18 \times 10^{10}\ M_\odot$ and $m_\mathrm{gas} = 2.08 \times 10^{9}\ M_\odot$, and the gravitational softening length is $\epsilon = 44\ \mathrm{kpc}$. A spherical overdensity halo finder is used to locate halos, where the mass $M_\mathrm{200m}$ within radius $R_\mathrm{200m}$ has an average density that is equal to 200 times the \textit{matter} density. The halo mass function is accurate to $\lesssim 5\%$ for halos with $\gtrsim 200$ (gas and dark matter) particles. The simulation was run on the NASA Endeavour supercomputer using 256 cores and taking only 10k CPU-hours.

%\newpage % formatting

%%%%%%%%%%%%%%%%%%%%%%%%%%%%%%%%%%%%%%%%%
\subsection{Galaxy Cluster Sample} 
\label{sec:clusters}
%%%%%%%%%%%%%%%%%%%%%%%%%%%%%%%%%%%%%%%%%

\begin{figure}[t]
\includegraphics[width=\linewidth]{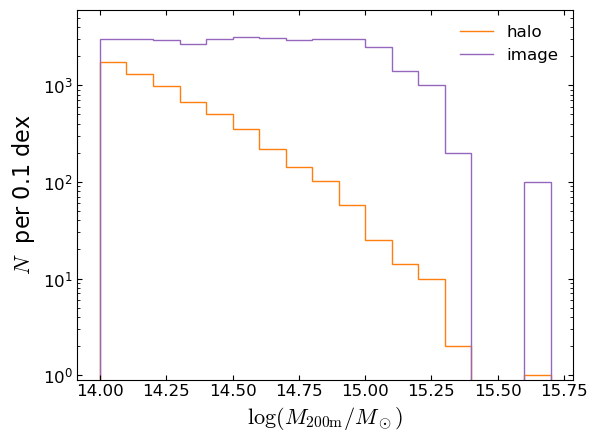}
\caption{The number of massive clusters (orange) decreases exponentially in the high-mass tail of the halo mass function. From these simulated clusters, we generate mock images (purple) to maintain approximately 3000 per 0.1 dex in mass except at the very highest masses, resulting in a uniform mass prior to avoid biases in training.}
\label{fig:counts}
\end{figure}

The $\texttt{HYPER}$ simulation described above produces 6,233 galaxy clusters at redshift $z=0$. The smallest halo has mass $M_\mathrm{200m} = 1.00 \times 10^{14}\ M_\odot$ and is resolved with $\sim 14000$ particles within a radius $R_\mathrm{200m} = 1.44\ \mathrm{Mpc}$. The largest halo has $M_\mathrm{200m} = 4.82 \times 10^{15}\ M_\odot$ and contains $\sim 700,000$ particles within $R_\mathrm{200m} = 5.21\ \mathrm{Mpc}$. Figure \ref{fig:counts} shows that the number of halos decreases exponentially in the high-mass tail of the distribution. To avoid biases in the ML model, we create a flat distribution with logarithmic mass \citep[e.g.][]{2015ApJ...803...50N, 2019ApJ...887...25H}. Overabundant low-mass halos are subsampled, while additional sightlines are generated for underabundant high-mass halos. Sightlines are sampled from a Fibonacci sphere \citep[e.g.][]{2010MatGe..42...49G} to ensure a more uniform and isotropic distribution. To generate each sightline, the particle coordinates are rotated so that the original $z$-axis aligns with the specified direction. Figure \ref{fig:counts} shows that the number of cluster images generated is flat with approximately 3000 per 0.1 dex in mass, except at the very highest masses where we impose a maximum number of sightlines per cluster.

For a given cluster sightline, we first construct three-dimensional fields, such as density and pressure, by mapping the particles onto a Cartesian mesh. The mesh length $l_\mathrm{mesh} = 12.8$ Mpc is slightly larger than the diameter of the largest halo, while the cell size $l_\mathrm{cell} = 100$ kpc is approximately twice the gravitational softening length and comparable to the effective resolution of the $\texttt{HYPER}$ simulation. Following the SPH approach, the density field for each particle species $\mathrm{x} = \mathrm{Gas, DM}$ is calculated using a weighted summation over nearby particles $j$ with mass $m_j$ and smoothing length $h_j$:
\begin{equation} \label{eq:density}
    \rho_\mathrm{x}(\boldsymbol{r}) = \sum_j m_j W(\boldsymbol{r} - \boldsymbol{r}_j, h_j) .
\end{equation}
The smoothing kernel is chosen to be a Wendland function \citep[e.g.][]{2012MNRAS.425.1068D},
\begin{equation}
    W(q,h) = \frac{21}{2\pi h^3}(1-q)^4(1 + 4q) ,
\end{equation}
where $q = r/h \in [0,1]$, consistent with the implementation used to compute densities for both gas and dark matter particles in the \texttt{HYPER} simulation. Each particle has an individual smoothing length that is adaptively calculated during the simulation. Note that this smoothing procedure differs from the gravitational softening, which uses a spline kernel with a constant softening length. Other fields such as the electron pressure (Section \ref{sec:sz}) and X-ray luminosity density (Section \ref{sec:xray}) are calculated similarly to Eq.~\ref{eq:density} \citep[see][for a review]{2010ARA&A..48..391S}.

For a given field, we then project along each of the three cardinal axes to make two-dimensional images with $N_\mathrm{pixel} = 128$ pixels per side. The image resolution, set by the pixel size $l_\mathrm{pixel} = 100\ \mathrm{kpc}$, is comparable to current X-ray observations and to followup SZ observations. For a given density field $\rho_\mathrm{x}(\boldsymbol{r})$, we make up to three projected density images with signal,
\begin{equation}
    S_\mathrm{x}[i,j] = \frac{\int \rho_\mathrm{x}(\boldsymbol{r}) dl}{\int \bar{\rho}_\mathrm{x} dl} = \frac{\sum_k \rho_\mathrm{x}[i,j,k]l_\mathrm{pixel}}{\sum_k \bar{\rho}_\mathrm{x} l_\mathrm{pixel}} ,
\end{equation}
by summing the density array $\rho_\mathrm{x}[i,j,k]$ along each of the three cardinal axes and multiplying by the pixel length. The projected density images are normalized using the cosmic average density $\bar{\rho}_\mathrm{x} = \Omega_\mathrm{x}\rho_\mathrm{crit}$, making them dimensionless.

%%%%%%%%%%%%%%%%%%%%%%%%%%%%%%%%%%%%%%%%%
\subsection{Sunyaev-Zel'dovich Effect}
\label{sec:sz}
%%%%%%%%%%%%%%%%%%%%%%%%%%%%%%%%%%%%%%%%%

The Sunyaev-Zel'dovich (SZ) effect arises when cosmic microwave background (CMB) photons are scattered by ionized electrons in the ICM and IGM \citep{1970CoASP...2...66S, 1972CoASP...4..173S}. Being the dominant secondary temperature anisotropy on arcminute scales, it is a promising probe of the growth of structure. Examples include direct detection of galaxy clusters, autocorrelation of temperature fluctuations, and cross-correlation with large-scale structure.

%The SZ effect is commonly considered to have two main components. The thermal SZ (TSZ) effect arises from inverse Compton scattering of the CMB with hot electrons, \rev{while the kinetic SZ (KSZ) effect is due to bulk motion of the gas relative to the CMB, resulting in a Doppler term due to scattering with fast electrons. Because the KSZ is subtle and probes the integrated electron momentum, it is less useful for this effort, and so we focus on the TSZ effect from the ICM for this paper.} 

We focus on the thermal SZ (tSZ) effect, which arises from inverse Compton scattering of the CMB with hot electrons. The CMB temperature fluctuations $\Delta T/T_\mathrm{cmb}$ are related to the Compton $y$ parameter multiplied by a frequency-dependent function. The Compton $y$ parameter is proportional to the integrated electron pressure along a given sightline,
\begin{equation}
    y = \frac{\sigma_\mathrm{T}}{m_\mathrm{e}c^2}\int P_\mathrm{e}(\boldsymbol{r}) dl .
\end{equation}
In the $\texttt{HYPER}$ simulation at $z= 0$, both the ICM and IGM are modeled as fully ionized, with one electron per hydrogen and two electrons per helium. A mean molecular weight $\mu_\mathrm{e} = 1.13$ is used when calculating the electron density or pressure. For a given electron pressure field $P_\mathrm{e}(\boldsymbol{r})$, we make up to three Compton $y$ images with signal,
\begin{equation}
    S_\mathrm{tsz}[i,j] = T_\mathrm{cmb}y[i,j] ,
\end{equation} 
by summing the pressure array along each of the three cardinal axes. While the Compton $y$ parameter is dimensionless, it is often multiplied by $T_\mathrm{cmb}$ to have convenient units of $\mu$K.

%%%%%%%%%%%%%%%%%%%%%%%%%%%%%%%%%%%%%%%%%
\subsection{X-ray Emission}
\label{sec:xray}
%%%%%%%%%%%%%%%%%%%%%%%%%%%%%%%%%%%%%%%%%

The ICM is a superheated plasma that radiates through thermal bremsstrahlung (free-free) emission. This emission can be detected at X-ray wavelengths using telescopes such as XMM-Newton, \textit{Chandra}, eROSITA. The apparent X-ray luminosity of a cluster is a probe of the temperature and mass of its ICM which, under the assumption of hydrostatic equilibrium, can be directly related to the total system mass \citep{ettori2019hydrostatic}. This ICM-based probe of cluster mass has been used broadly in X-ray surveys to characterize both individual systems \citep[e.g.][]{allen2002chandra,aghanim2011planck, ilani2024galaxy} and general population statistics \citep[e.g.][]{vikhlinin2009chandra, lovisari2017x, bahar2022erosita}.

For a plasma of ionized hydrogen, helium, and metals, the bolometric emissivity or luminosity density is given by
\begin{equation}
    l_\mathrm{bol} \approx 1.4\times 10^{-27}g_\mathrm{ff}T^{1/2}n_\mathrm{e}\sum_i z_i^2n_i \ \mathrm{erg\,s^{-1}\,cm^{-3}} ,
    \label{eq:xry}
\end{equation}
where $g_\mathrm{ff}$ is the free-free gaunt factor, $T$ is the plasma temperature, $n_\mathrm{e}$ is the electron number density, and $z_i$
and $n_i$ are the charge and number density of the various ions $i$ \citep[see][for a review]{1988S&T....76Q.639S}. In the $\texttt{HYPER}$ simulation, the ICM is modeled as a fully ionized plasma with an average metallicity of $0.3 Z_\odot$ at $z=0$. For a given luminosity density field $l_\mathrm{bol}(\boldsymbol{r})$, we make up to three X-ray surface brightness images with signal,
\begin{equation}
    S_\mathrm{xry}[i,j] = \int l_\mathrm{bol}(\boldsymbol{r}) dl = \sum_k l_\mathrm{bol}[i,j,k]l_\mathrm{pixel},
    \label{eq:sz}
\end{equation}
by summing the luminosity density array $l_\mathrm{bol}[i,j,k]$ along each of the three cardinal axes. The surface brightness images have units of erg$\,$s$^{-1}\,$cm$^{-2}$.

For this pilot study, we omit noise from the mock SZ and X-ray images to assess the baseline performance of the method. Realistic noise models will be incorporated in future work before applying the approach to observational data.
\section{Machine Learning} \label{sec:machine_learning}

Our deep learning pipeline consists of two stages. First, we pre-process the mock observations and split our datasets into training and testing portions (Section \ref{sec:prep}). Second, we train the score model on the training dataset to reconstruct the gas and dark matter density maps conditioned on the SZ and X-ray inputs. In Section \ref{sec:score_models}, we describe the framework of stochastic differential equations and how score models utilize Langevin dynamics sampling to estimate the data distribution.

%%%%%%%%%%%%%%%%%%%%%%%%%%%%%%%%%%%%%%%%%
\subsection{Dataset Construction}
\label{sec:prep}
%%%%%%%%%%%%%%%%%%%%%%%%%%%%%%%%%%%%%%%%%

\begin{figure}
    \begin{center}
        \includegraphics[width=1\linewidth]{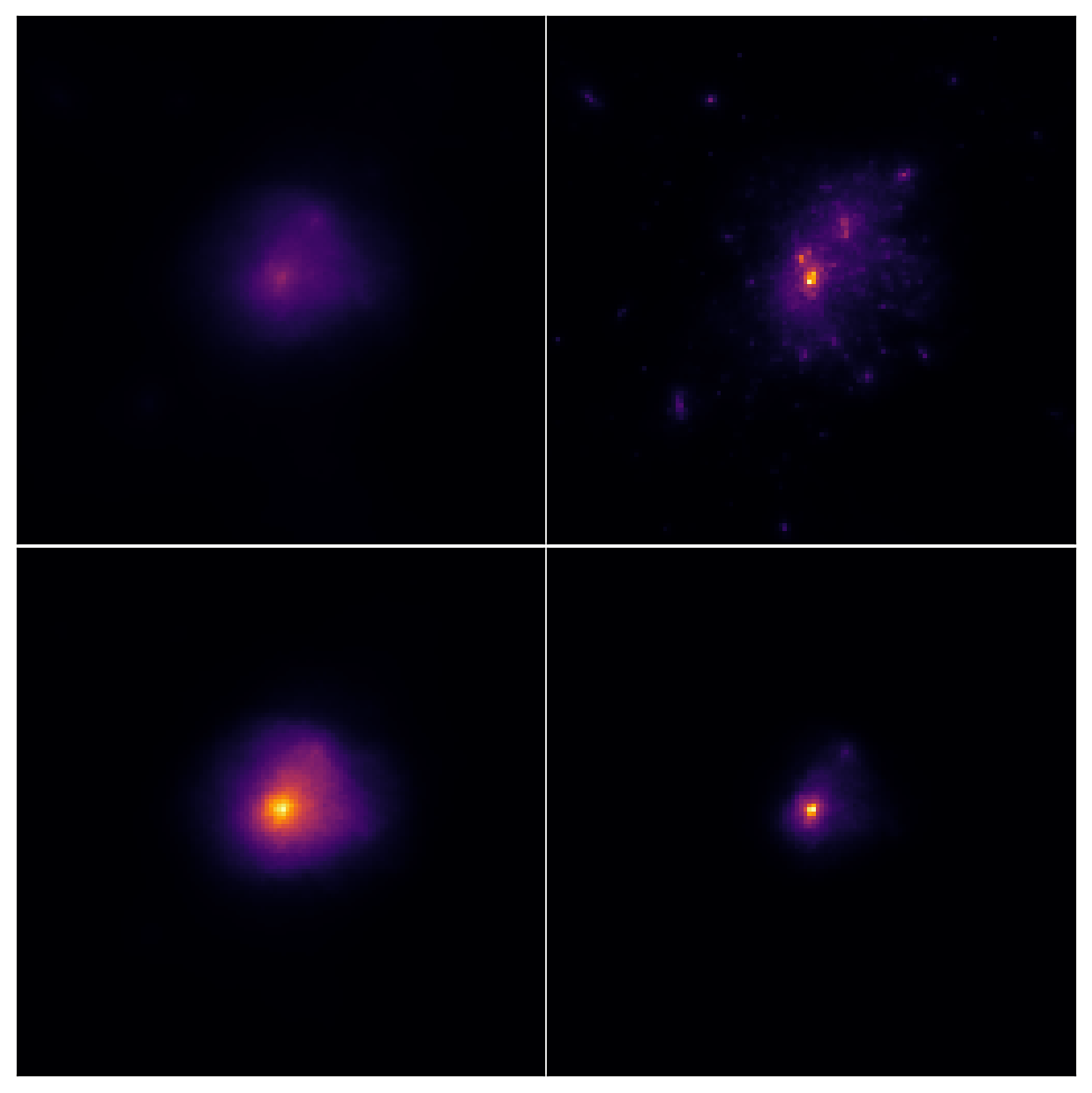}
    \end{center}
    \caption{Simulated mass maps of gas (top left), dark matter (top right), and mock observations of SZ (bottom left), and X-ray (bottom right), of a large mass halo. The gas and dark matter are underlying distributions that we are trying to compute, while the SZ and X-ray are the observables.}
    \label{fig:sim_data}
\end{figure}

Figure~\ref{fig:sim_data} shows sample simulation maps of gas and dark matter, and mock observations of SZ and X-ray maps of a cluster. All images are 128 by 128 pixels, with 0.1 Mpc/pixel or 12.8 Mpc in physical length. We normalize these images to within 0 and 1 using global upper and lower bounds for each type of image since we want to preserve the relative intensities between clusters of different sizes: we want our model to generate the mass maps with intensities associated with the input galaxy cluster size.

Our dataset consists of 34,714 sets of (SZ, X-ray, gas, dark matter) maps of galaxy clusters, which is then split into training, validation, and testing portions. We will only train our model using the training set and improve on any hyperparameters using the validation set, reserving the testing set for an unbiased evaluation of our model after the full training pipeline. In addition, while some galaxy clusters may have multiple projections in our full dataset, we ensure that all projections of each galaxy cluster are contained in exactly one of the training, validation, and testing sets. This ensures that the model does not train on any of the clusters in the test set. 

%%%%%%%%%%%%%%%%%%%%%%%%%%%%%%%%%%%%%%%%%
\subsection{Conditional Score-based Models} 
\label{sec:score_models}
%%%%%%%%%%%%%%%%%%%%%%%%%%%%%%%%%%%%%%%%%

\begin{figure*}
    \begin{center}
        \includegraphics[width=1\linewidth]{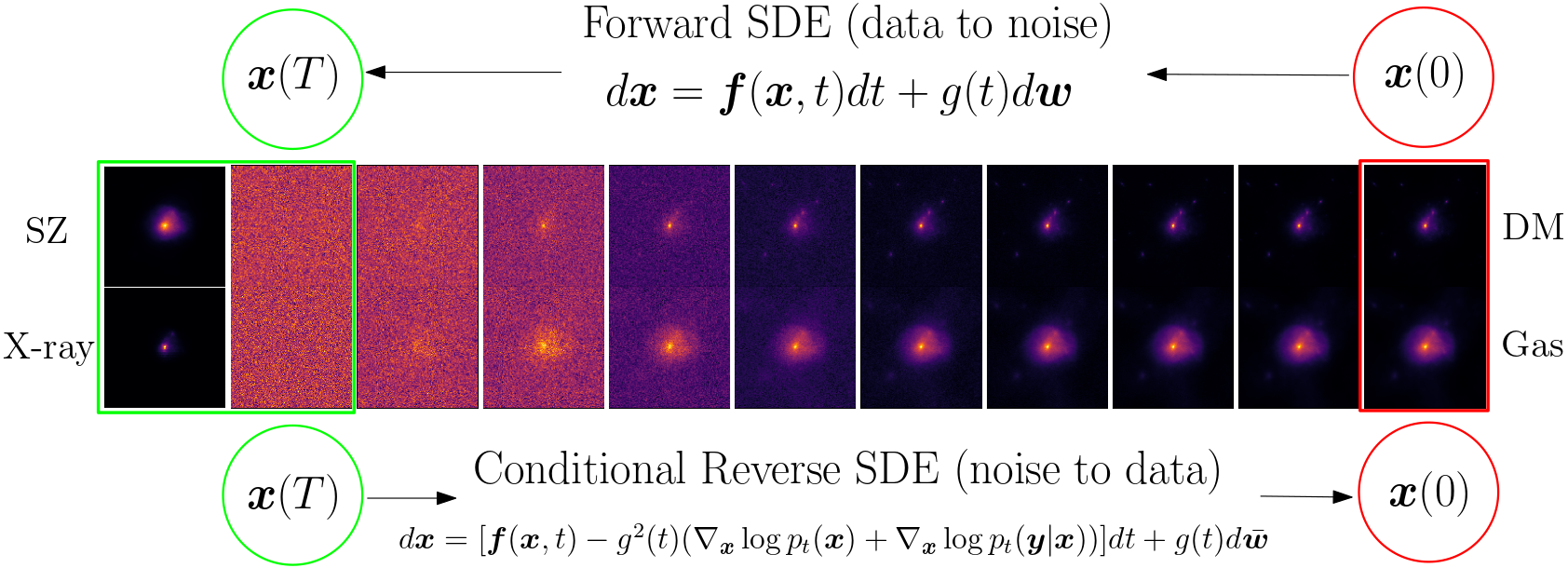}
    \end{center}
    \caption{Pedagogical diagram of sampling dark matter (top row) and gas (bottom row) maps using the Euler-Maruyama discretization of Stochastic Differential Equations (SDE), as previously shown in Figure 1 in \cite{Song_2021}. Before training, we define the forward SDE (top equation), which transforms the data distribution to a random distribution by successively adding noise to the images. During training, the model learns the score of the data distribution in order to sample from it using the conditional reverse SDE (bottom equation), which transforms the random distribution back to the data distribution conditioned on the SZ and X-ray observations (left column). The model utilizes the conditional reverse SDE to generate different realizations of the gas and dark matter maps based on the initial sampled Gaussian random field. The SZ and X-ray inputs along with the initial random samples for the gas and dark matter (green) represent the inputs to our image-to-image diffusion model, while the predicted gas and dark matter (red) represent the outputs.}
    \label{fig:sde}
\end{figure*}

Our score model transforms a sample from a Gaussian random field into a realization of the gas and dark matter map of a galaxy cluster, which effectively samples from the gas and dark matter posterior distribution. Figure~\ref{fig:sde} describes the sampling process of a score-based model using stochastic differential equations (SDEs) \citep[e.g.][]{Song_2021}. We provide the forward SDE to gradually transform the gas and dark matter maps into random noise, and the diffusion model learns the reverse SDE that maps the noise back to the gas and dark matter maps. In particular, the model learns to sample from the joint posterior of gas and dark matter maps of clusters conditioned on the SZ and X-ray inputs. This section discusses the statistical framework behind the sampling techniques for the posterior, and the training process for a score-based diffusion model.

We first provide a forward mapping from the final data distribution (gas and dark matter distribution) to a prior (noise distribution) by continuously injecting increasing noise using a forward-time It\^o stochastic differential equation (SDE),
\begin{equation}
    d\bm{x} = \bm{f}(\bm{x}(t),t)dt + g(t)d\bm{w},
    \label{eq:sde_forward}
\end{equation}
also shown at the top of Figure~\ref{fig:sde}. Here, $\bm{x} = (\text{Gas}, \text{DM})$ are the gas and dark matter maps, which are a function of time $t$ that defines the noise level of the process. $\bm{f}$ is the drift component, which describes the deterministic force that pushes on $\bm{x}$: for our case, this will be 0, as $\bm{x}$ eventually evolves into purely random noise. $g$ is the diffusion component, which describes the scale of noise added to the system. $\bm{w}$ is the standard Wiener process, or Brownian motion. The forward SDE maps $\bm{x}(t=0)$ sampled from the data distribution to $\bm{x}(t=T)$ sampled from a Gaussian random field by stochastically integrating Equation \ref{eq:sde_forward} using standard SDE methods such as Euler-Maruyama \citep[e.g.][]{Maruyama_1955, Song_2021}. 

The model learns to sample the data distribution $p_t(\bm{x})$ by solving the reverse-time SDE \citep[e.g.][]{Anderson_1982} 
\begin{equation}
    d\bm{x} = [\bm{f}(\bm{x}(t),t) - g^2(t)\nabla_{\bm{x}}\log p_t(\bm{x})]dt + g(t)d\bm{\bar{w}},
    \label{eq:sde_reverse}
\end{equation}
which inverts the forward-time transformation by mapping $\bm{x}(t=T)$ to $\bm{x}(t=0)$, also shown at the bottom of Figure~\ref{fig:sde}. This equation requires knowing the score of the data distribution $\nabla_{\bm{x}}\log p_t(\bm{x})$, or the gradient with respect to the data of the log probability density at some noise level $\sigma(t)$. We estimate the score using machine learning by parameterizing it with a deep network $s_{\theta}(\bm{x}(t), t)$.

Learning the distribution of $\nabla_{\bm{x}}\log p_t(\bm{x})$ directly is complex, since we do not have access to the true data distribution $p(\bm{x})$, and thus we can only reply on supervised signals derived from known conditionals: \cite{Song_2021} uses a technique that introduces noise to the data distribution, and solves for the conditional noisy distribution instead. For some parameterization of the noise $\sigma(t)$, we can reparameterize $\bm{x}(t)$ as a noisy data point $\bm{\tilde{x}} = \bm{x} + \sigma\xi$, where $\xi \sim N(0, 1)$ is a normal distribution or any known noise distribution. Then, $p(\bm{\tilde{x}} | \bm{x})$ is analytically a Gaussian, and thus the score is also analytically known:
\begin{equation}
    \nabla_{\bm{\tilde{x}}} \log p_{\sigma}(\bm{\tilde{x}}| \bm{x}) = -\frac{\bm{\tilde{x}} - \bm{x}}{\sigma^2}.
    \label{eq:cond_score}
\end{equation}
Using a key result from the denoising score matching procedure \citep[e.g.][]{Vincent_2011, Song_2021}, optimizing the score matching relative to $\nabla_{\bm{\tilde{x}}} \log p_{\sigma}(\bm{\tilde{x}}| \bm{x})$ implicitly optimizes the score matching to the true marginal score $\nabla_{\bm{\tilde{x}}}\log p_\sigma(\bm{\tilde{x}})$ up to a constant. In other words, instead of solving a very hard optimization problem that uses the objective 
\begin{equation}
    \underset{p_\sigma(\bm{\tilde{x}})}{\mathbb{E}} \norm{s_{\theta}(\bm{\tilde{x}}, \sigma) - \nabla_{\bm{\tilde{x}}}\log p_\sigma(\bm{\tilde{x}})}_2^2,
\end{equation}
we utilize a known form of the conditional score in equation \ref{eq:cond_score} and optimize the alternative objective
\begin{equation}
    \underset{p(\bm{x})p(\bm{\tilde{x}} | \bm{x})}{\mathbb{E}} \norm{s_{\theta}(\bm{\tilde{x}}, \sigma) + \frac{\bm{\tilde{x}} - \bm{x}}{\sigma^2}}_2^2,
    \label{eq:surrogate_obj}
\end{equation}
which effectively minimizes the normalized difference between the noised and unnoised maps. This method cleverly utilizes the conditional distribution only during training as a surrogate to optimize the marginal score that we want to estimate. In practice, the expectation is approximated by the samples of our data, and the optimization process produces the best $\theta^*$ that best matches the score. The trained model $s_{\theta^*}(\bm{\tilde{x}}, \sigma)$ is then used as an estimate of the score term $\nabla_{\bm{x}}\log p_t(\bm{x})$ in equation \ref{eq:sde_reverse}. Note that $\sigma$ and $t$ are used interchangeably here, and in particular we describe the noising process in SDEs with the independent variable $t$, but during score matching we describe the noise as $\sigma$.

Next, a model trained to learn the score from the objective function \ref{eq:surrogate_obj} will be able to sample realizations of dark matter and gas maps from a Gaussian random field. However, in our framework we would like to not only learn the distribution of dark matter and gas, but also be able to sample from it conditioned on SZ and X-ray maps, that is, we would like to learn the posterior $p_t(\bm{x} | \bm{y})$, where $\bm{y} = (\text{SZ}, \text{X-ray})$ are the conditional inputs. We can rewrite the reverse SDE to learn a mapping from $p_{t=T}(\bm{x} | \bm{y})$ to $p_{t=0}(\bm{x} | \bm{y})$, given by
\begin{equation}
\begin{split}
    d\bm{x} &= [\bm{f}(\bm{x},t) - g^2(t)\nabla_{\bm{x}}\log p_t(\bm{x}|\bm{y})]dt + g(t)d\bm{\bar{w}}.
\end{split}
\label{eq:sde_cond_reverse}
\end{equation}
In the case where we analytically understand the likelihood of our samples $p(\bm{y} | \bm{x})$, we would use Bayes rule to separate the prior from the likelihood in the score, namely $\nabla_{\bm{x}}\log p_t(\bm{x}|\bm{y}) = \nabla_{\bm{x}}\log p_t(\bm{x}) + \nabla_{\bm{x}}\log p(\bm{y}|\bm{x})$, and only use machine learning to estimate the first term and analytically evaluate the second. However, since it is difficult to express the relationship between the SZ and X-ray maps to the Gas and DM maps analytically, we will parameterize the full score $s_\theta(\bm{\tilde{x}}, t, \bm{y})$ using a model that takes in $\bm{y}$ as additional channels. The corresponding objective is then modified to perform an expectation over all $(\bm{x}, \bm{y})$ data pairs:
\begin{equation}
    \underset{p(\bm{x, y})p(\bm{\tilde{x}} | \bm{x})}{\mathbb{E}} \norm{s_{\theta}(\bm{\tilde{x}}, \sigma, \bm{y}) + \frac{\bm{\tilde{x}} - \bm{x}}{\sigma^2}}_2^2,
\end{equation}
and once trained this conditional network is likewise substituted into the score term in equation \ref{eq:sde_cond_reverse}.

Finally, the noising procedure that we will use is described by the exponential schedule
\begin{equation}
    \sigma(t) = \sigma_{min}*\left[\frac{\sigma_{max}}{\sigma_{min}}\right]^{t/T},
    \label{eq:noising}
\end{equation}
where the variance at time $t = 0$ is $\sigma_{min}$, defined to be the scale of noise in the data distribution, and the variance at time $t = T$ is $\sigma_{max}$, which sets the scale of the noise in the Gaussian random field. In general, we want to set $\sigma_{min}$ to be a reasonable lower bound for the noise we find and want to reconstruct in the data, and $\sigma_{max}$ to be comparable to the normalized pixel values of the images. A balance is to be found for the range of these variance bounds: a noising procedure with a small change in $\sigma$ may not learn the small-scale structures and desired accuracy in the data distribution, while a large change in $\sigma$ may not be able to learn the data distribution at all.

Our diffusion model uses the Noise Conditional Score Network (NCSN++) architecture and a Variance-Exploding SDE sampling process \citep[VESDE, as described in appendix B of][]{Song_2021} to learn the conditional score of the distribution. The NSCN++ is a variant of RefineNet \citep[][]{Lin_2016}, a U-Net variant with instance normalization and dilated convolutions \citep[c.f.~][appendix A]{Song_2019}. For training, the model has 2 residual blocks per resolution, $\sigma_{max} = 1$, and $\sigma_{min} = 0.001$. For sampling, the variance-exploding procedure has a drift term of $\bm{f} = 0$ and a diffusion term $g = \sqrt{d\sigma^2/dt}$. We use the implementation given by Alexendre Adam's repository \citep[e.g.][]{Adam_2022, Legin_2023} that can be found at this link: \href{https://github.com/AlexandreAdam/score_models}{https://github.com/AlexandreAdam/score\_models}.

\section{Results} \label{sec:results}

To validate generated results from our score model, we apply a series of performance metrics to our model with an independent testing set of mock observations. We present sample generated gas and dark matter maps along with their corresponding sampling convergence (Section \ref{sec:imgs}). We then compute the radial density profiles and pixel-wise fractional error histograms (Section \ref{sec:rad}) and bias and cross correlation coefficients (Section \ref{sec:corr}), and demonstrate that our model is able to generate maps that reconstruct these quantities with high accuracy.

\begin{figure}[htbp]
    \begin{center}
        \includegraphics[width=1\linewidth]{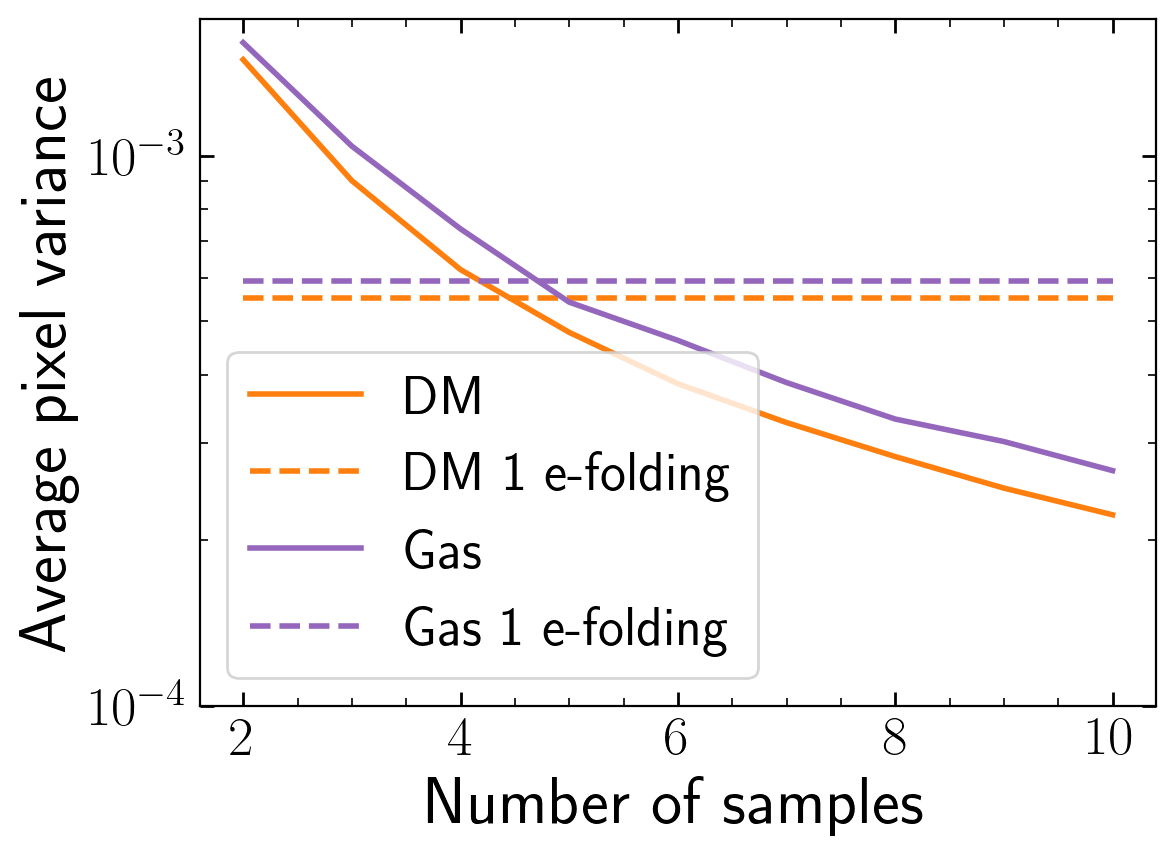}
    \end{center}
    \caption{Sample variance of mean pixel gas predictions (purple) and dark matter (orange) as a function of the number of samples. Since each starting sample is a different noise realization, the output gas and dark matter maps will be slightly different as well: the sample variance is hence a way of assessing whether the mean of these output images converged. As we increase the number of samples, the variance decreases as we converge to the mean value of the posterior distribution: at around five samples, the variance undergoes one e-folding of a two-sample variance (dotted lines). We thus select ten samples as a safe threshold when computing our average predictions.}
    \label{fig:sample_var}
\end{figure}

\subsection{Mass Maps}

\label{sec:imgs}

\begin{figure*}[htbp]
    \begin{center}
        \includegraphics[width=1\linewidth]{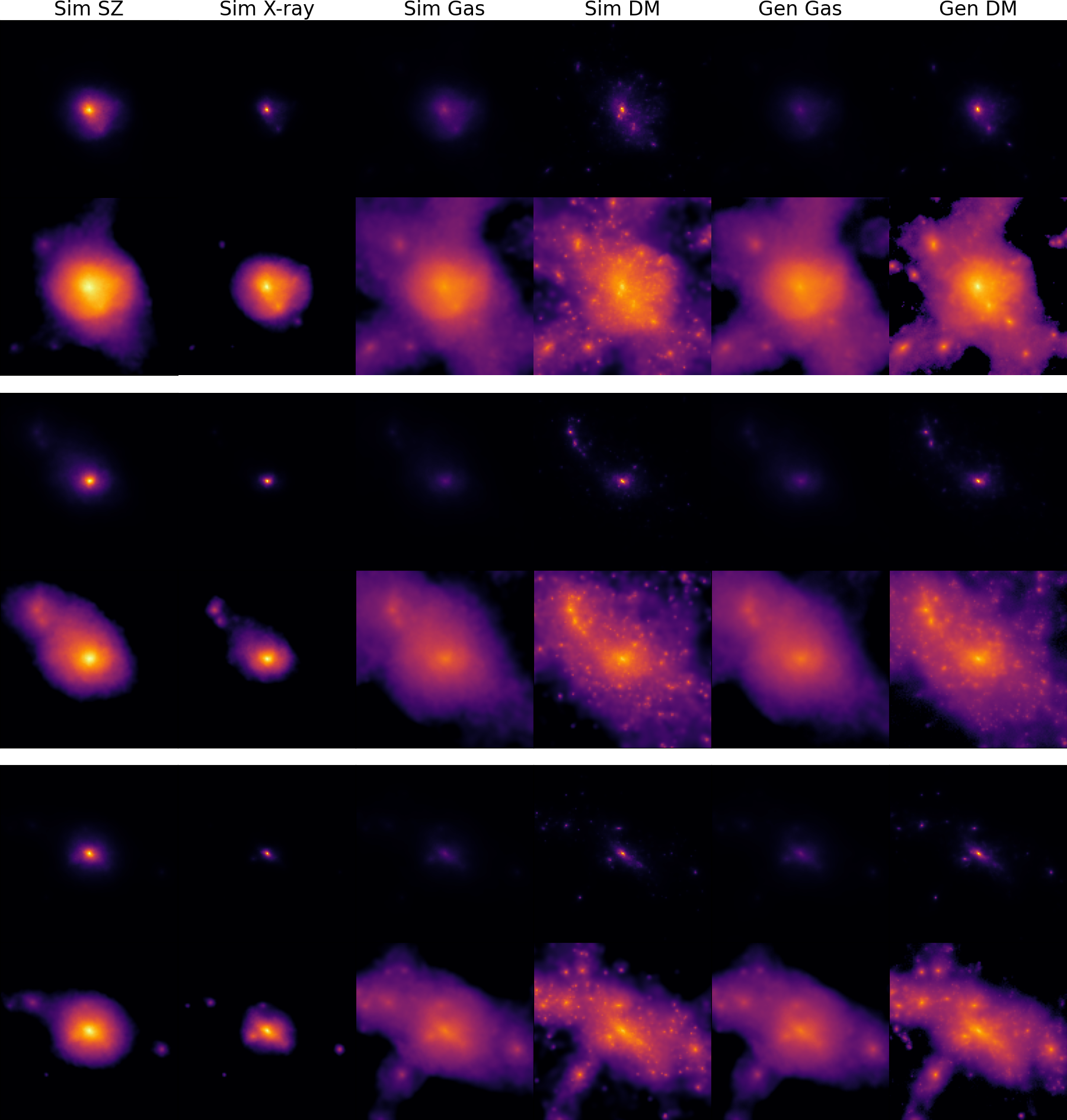}
    \end{center}
    \caption{Selected model predictions of gas and dark matter maps (rightmost 2 columns) conditioned on SZ and X-ray inputs (leftmost 2 columns) along with ground truth gas and dark matter maps (middle 2 columns). For each test cluster, we show the linear maps on the top row and the log-transformed maps on the bottom row. The predicted maps are averages of 10 samples from the learned joint posterior. While our models are trained on data in linear scale, the images are also displayed in logarithmic scale to enhance dynamic range and highlight substructure. Across large and small mass clusters, we see accurate reconstruction for large and small-scale structures, particularly noticeable when we log-transform the generated images.}
    \label{fig:score_results}
\end{figure*}

\begin{figure*}
    \begin{center}
        \includegraphics[width=1\linewidth]{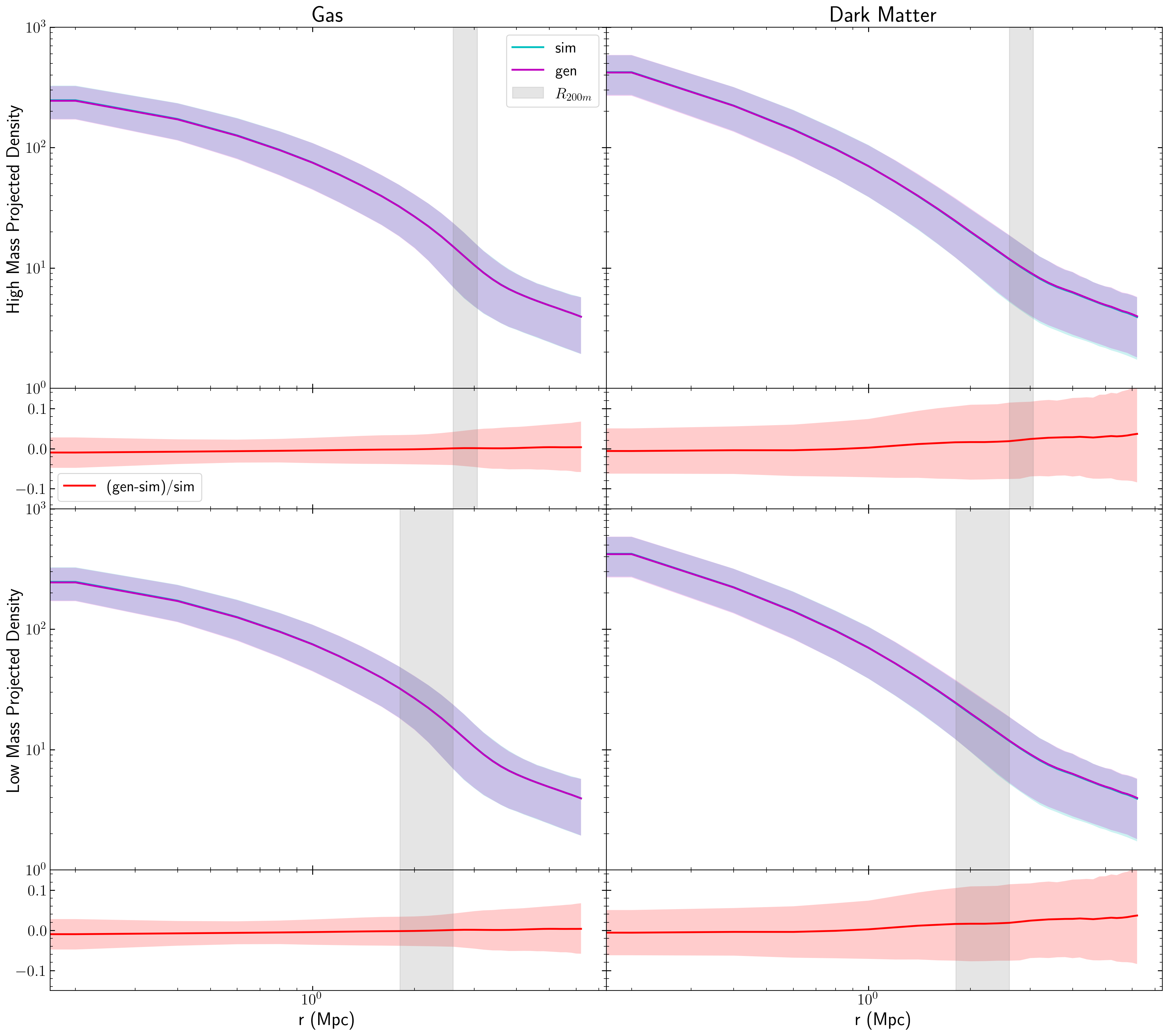}
    \end{center}
    \caption{Radial density profiles for dark matter (left) and gas (right). Profiles of high mass clusters ($6 \times 10^{14} \le M_{200m} / M_\odot \le 1 \times 10^{15}$) are shown on the top, and those of low mass clusters ($2 \times 10^{14} \le M_{200m} / M_\odot < 6 \times 10^{14}$) are shown on the bottom. In each panel we graph the generated maps (magenta) and simulated maps (cyan) with the 16th to 84th percentiles shaded. The ranges for $R_{200m}$ (shaded gray) in Mpc are $2.60 \le R_{200m} \le 3.06$ for large clusters and $1.81 \le R_{200m} \le 2.60$ for small clusters. Under each curve, we show the fractional difference between the generated and simulated maps in red. For the gas, we see agreement to within 5\% for all scales for both high and low mass clusters, with a slightly larger fractional error for $r > 1$ Mpc. For the dark matter, the fractional error spread increases to around 10\% at the smallest spatial scales.}
    \label{fig:density}
\end{figure*}

The diffusion model is trained on clusters of sizes $1.00 \times 10^{14} \le M_{200m} / M_\odot \le 4.82 \times 10^{15}$, but we will only test our model on clusters within the range $2.00 \times 10^{14} \le M_{200m} / M_\odot \le 1.00 \times 10^{15}$. Previous deep-learning based estimates of cluster masses \citep[e.g.][]{2019ApJ...887...25H, Ntampaka_2019, yan2020galaxy} broadly observed a `mean-reversion' effect, wherein the masses of small clusters were overpredicted and those of large clusters were underpredicted. Such effects can be resolved with more training data and broader mass priors. As a result, we exclude these extremities and reserve their analysis for future work. For our test mass region, we will denote clusters within $2.00 \times 10^{14} \le M_{200m} / M_\odot < 6.00 \times 10^{14}$ as low mass clusters and $6.00 \times 10^{14} \le M_{200m} / M_\odot \le 1.00 \times 10^{15}$ as high mass clusters, roughly divided evenly in log mass space. The ranges for $R_{200m}$ in Mpc are $2.60 \le R_{200m} \le 3.06$ for large clusters and $1.81 \le R_{200m} \le 2.60$ for small clusters.

The diffusion model samples from the learned posterior by first sampling from a Gaussian random field, and mapping that to a realization of gas and dark matter densities using the reverse SDE and the specified noising schedule. Since the sampling process is probabilistic, the sample variance is used to determine the number of samples we need from the posterior to produce a converged mean prediction. In Figure~\ref{fig:sample_var}, we compute the sample variance averaged over all pixels as a function of the number of samples, and we observe that one e-folding of a two-sample variance occurs at around five samples for both the gas and dark matter. We will use ten samples, each of which is generated from a random Gaussian random field, to compute mean predictions in the following results. The samples are generated via the noising equation \ref{eq:noising} with $1000$ discrete time steps.

Figure~\ref{fig:score_results} shows three sample galaxy cluster predictions from our diffusion model, ranging across our full test mass range. Each pair of rows is a test cluster, with linear maps on the top row and the log-transformed maps on the bottom row. For the log-transformed maps, the dark matter and gas densities are clipped to a mean density of 1 before we apply the log-transform, and the SZ and X-ray inputs are clipped to a corresponding reasonable lower bound such that the log-transform shows the major substructures. While our models are trained on data in linear scale, the images are also displayed in logarithmic scale to enhance dynamic range and highlight substructure. The first 4 columns represent the ground truth simulation maps of SZ, X-ray, gas, and dark matter, respectively. Multiple realizations of the gas and dark matter maps are generated to compute statistical quantities: the final 2 columns represent the generated 10-sample average of the gas and dark matter maps conditioned on the SZ and X-ray inputs. Visually, our model is able to pick up on both the large-scale and small-scale features of the clusters across different mass bins (rows), which are especially noticeable in the log-transformed images. While the substructure signals in the SZ and X-ray may not be large in the simulation images, they are still present, and we find that the machine learning model is able to pick on these small signals in those regions and learn a complex internal model of the galaxy cluster to produce the substructures in the output gas and dark matter images. 

\subsection{Density Profiles}
\label{sec:rad}

\begin{figure}[htbp]
    \begin{center}
        \includegraphics[width=1\linewidth]{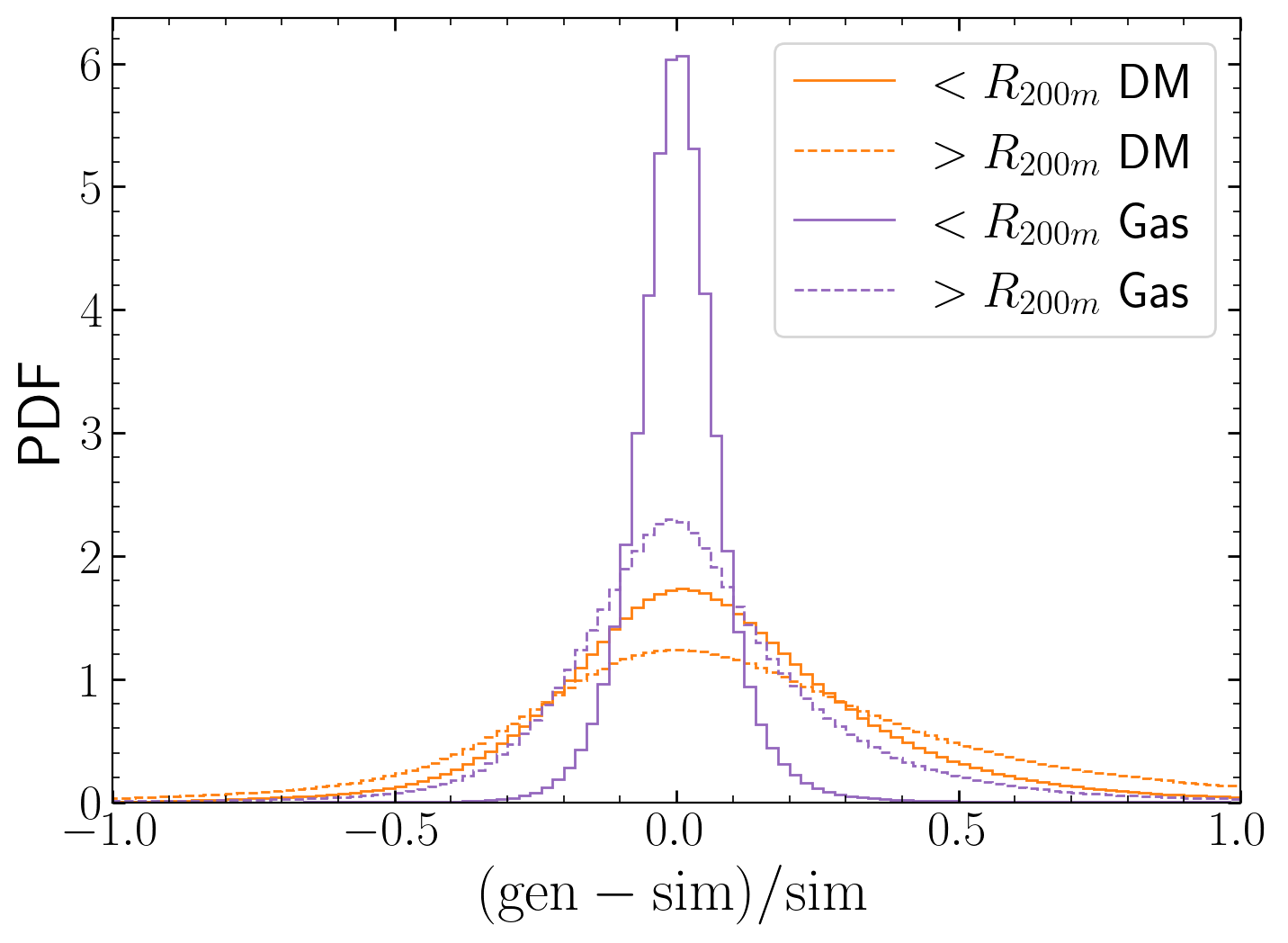}
    \end{center}
    \caption{Fractional error distribution for dark matter (orange) and gas (purple), for pixels inside $R_{200m}$ (solid line) and outside $R_{200m}$ (dashed line). We see a smaller spread in error for pixels inside $R_{200m}$, which indicates that our model reconstructs pixels within $R_{200m}$ better than those outside. This matches the error curve in the radial profiles shown in Figure~\ref{fig:density}, where we see low error near the center that increases in both bias and variance as we increase the radius.}
    \label{fig:frac_error}
\end{figure}

To validate the performance of mass reconstruction radially, we compute the average density profile for both the generated and simulated gas and dark matter maps, binning the pixels radially. Figure~\ref{fig:density} shows these profiles on the top panel, along with the associated fractional reconstruction error in the bottom panel. The simulation average radial profiles are shown in magneta, and the generated average radial profiles are shown in cyan, with the ranges of $R_{200m}$ shaded vertically in gray. Below each plot we show the fractional error in red. Overall we match both the mean and the 68\% spread in our reconstructed mass maps with the ground truth mass distribution, with a 1-2\% average fractional error across all clusters in the central $r < 1$ Mpc region, with a slightly larger fractional error at the edge of the cluster in regions $r \ge 1 $ Mpc. We also find that the standard deviation of the fractional error is the same order of magnitude as the bias. Matching both the mean trend and spread is equally important, as there is natural scatter from the simulation profiles from the various cluster sizes: the fact that our model is able to reproduce this uncertainty is a strong indication that it is able to distinguish between clusters of different sizes. 

We further demonstrate that our model tries to preserve the overall mass: the average total mass fractional error for the gas maps is $0.0023^{+0.0010}_{-0.0018}$, while that of the dark matter is $-0.0079^{+0.0070}_{-0.0044}$. Overall, we find that for both gas and dark matter, the reconstructed large clusters have a slight under-bias at $r > 1$ Mpc where the model tries to match the zero boundary condition near the edges of the maps. For small clusters, the model slightly overpredicts near the edges, where increase in the magnitude of the error beyond the virial radius is an artifact of the denominator in the fractional error: because there is minimal mass near the edge, the residual (numerator) becomes much larger than the denominator. Finally, in Figure~\ref{fig:frac_error} we show the pixel-wise fractional error distribution of the reconstructed images, split into pixels within and outside of $R_{200m}$. We see that the model more accurately reconstructs pixels within $R_{200m}$, while those outside $R_{200m}$ have a much flatter fractional error distribution.

\begin{figure*}[htbp]
    \begin{center}
        \includegraphics[width=1\linewidth]{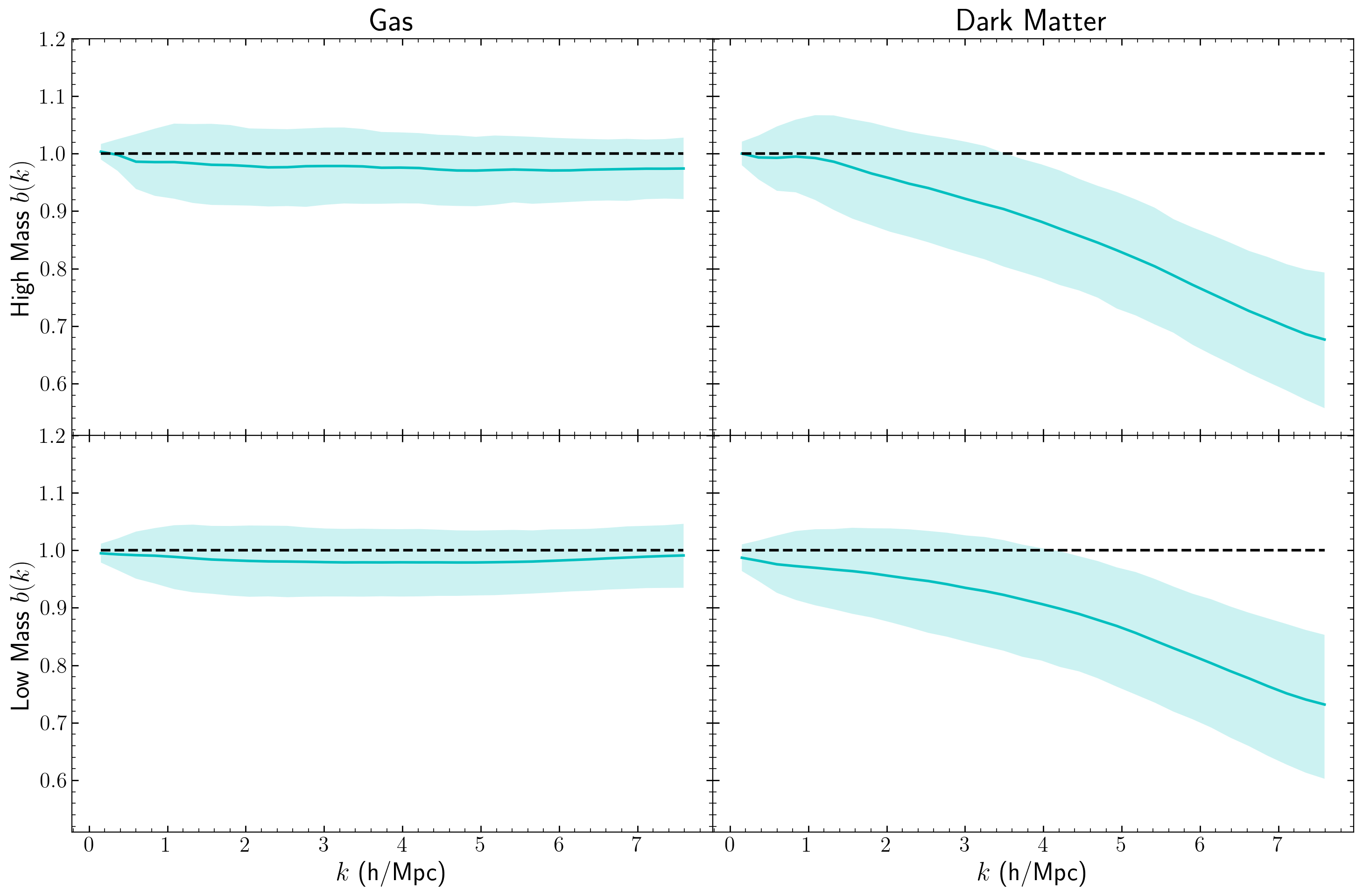}
    \end{center}
    \caption{Bias coefficients (blue) between the simulation and generated outputs (Equation~\ref{eq:bias}). Coefficients for high mass clusters ($6 \times 10^{14} \le M_{200m} / M_\odot \le 1 \times 10^{15}$) are shown on the top, and those of low mass clusters ($2 \times 10^{14} \le M_{200m} / M_\odot < 6 \times 10^{14}$) are shown on the bottom. The model achieves close-to-unity bias (black) across all scales.}
    \label{fig:bias}
\end{figure*}

Our results improve on the error of those of Table 4 in~\cite{Andres_2024}, where they compute the average total mass fractional error of generated density maps from SZ, X-ray and star observations using a U-Net model to be $-0.015^{+0.028}_{-0.026}$ for clusters of size $14.02 \le \log M_{200} / h^{-1}M_{\odot} < 14.63$, and $-0.024^{+0.019}_{-0.017}$ for clusters of size $14.63 \le \log M_{200} / h^{-1}M_{\odot} < 15.38$. Our model leverages the sampling power of diffusion models to converge to a much more accurate gas and dark matter map rather than a single point estimate as used in other image-to-image models.

\subsection{Cross Correlations}\label{sec:corr}

In addition to analyzing the spatial reconstruction of our models, we also analyze the accuracy of our results using the bias and cross correlation coefficients in Fourier space. First, the bias is the tendency to under-predict or over-predict pixel intensities in our models, given by the auto-correlation of the simulated maps $P_\mathrm{sim}$ and the generated maps $P_\mathrm{gen}$: 
\begin{equation}
    b(k) = \sqrt{\frac{P_\mathrm{gen}(k)}{P_\mathrm{sim}(k)}}.
\label{eq:bias}
\end{equation}
The bias $b(k)$ of a particular channel output is unbiased if $b=1$, overbiased if $b>1$, and underbiased if $b<1$. 

\begin{figure*}
    \begin{center}
        \includegraphics[width=1\linewidth]{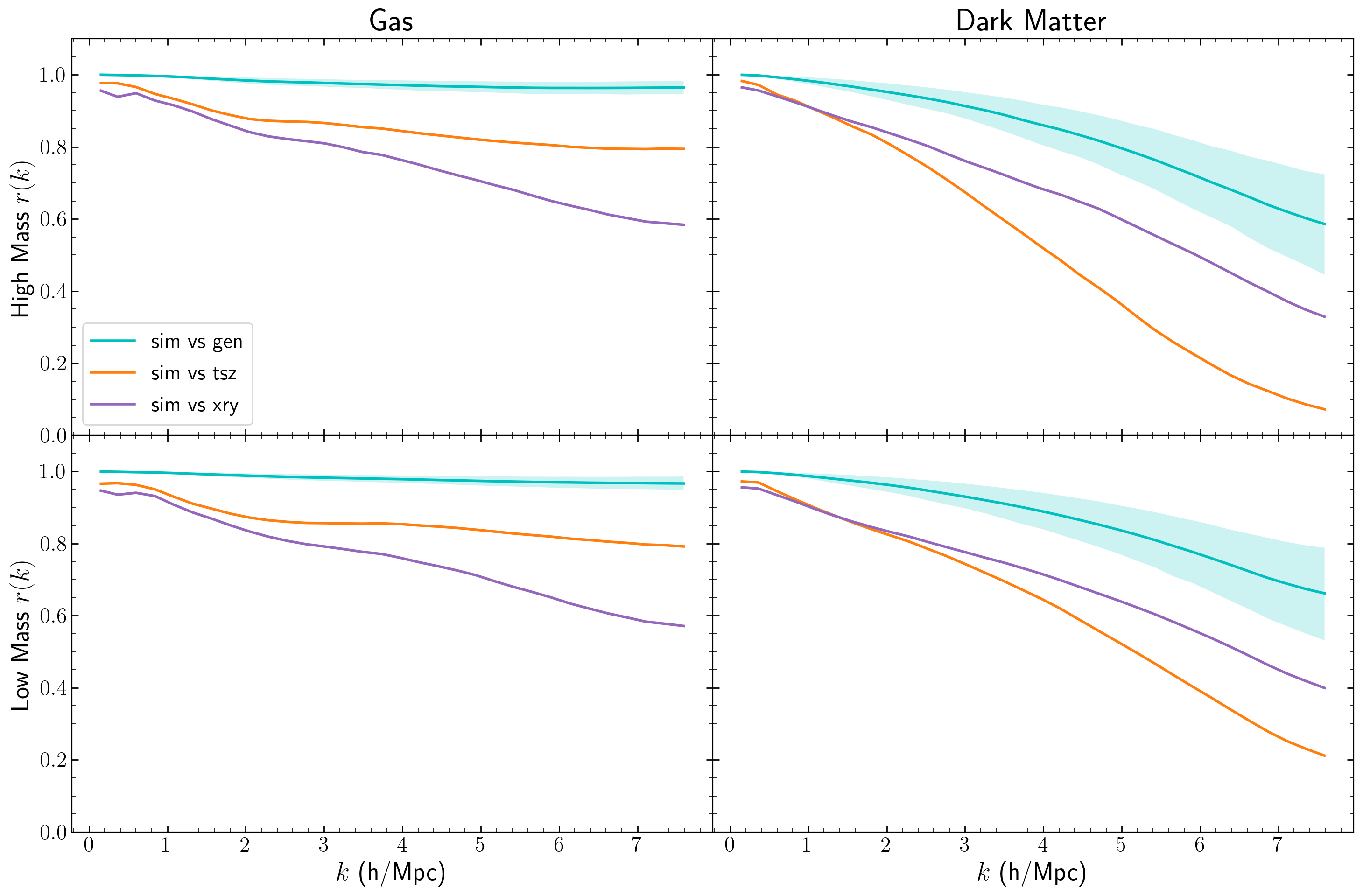}
    \end{center}
    \caption{Normalized cross-correlation coefficients between the simulation and generated outputs (blue) of gas (left) and dark matter (right). Coefficients for high mass clusters ($6 \times 10^{14} \le M_{200m} / M_\odot \le 1 \times 10^{15}$) are shown on the top, and those of low mass clusters ($2 \times 10^{14} \le M_{200m} / M_\odot < 6 \times 10^{14}$) are shown on the bottom. We expect near-perfect agreement of unity on large scales (small $k$), and a monotonically decreasing behavior as we move to smaller scales (large $k$). By comparing with the cross-coefficients between the simulated outputs and SZ (orange) and X-ray (purple), we see that our model is able to learn a much stronger nonlinear correlation than that of a scale-dependent linear bias model.}
    \label{fig:cc}
\end{figure*}

Figure~\ref{fig:bias} shows the bias coefficients for the simulated outputs and the generated outputs binned radially, assuming angular independence, averaged over all clusters. We see a similar behavior for predicting both high-mass and low-mass clusters. Because our model is learning from mock SZ and X-Ray inputs, which are derived from the simulated gas density, the bias for the generated gas maps is very close to one. For dark matter maps, there is a slight under-bias for small scales, likely because the signal from SZ and X-ray inputs do not fully capture the small-scale structures of the underlying dark matter. In addition, because the gas maps are more smooth in nature, the continuum between large and small scales should be easier to predict compared to dark matter maps.

Second, by computing the normalized cross-correlation coefficients, we can analyze whether or not the generated outputs are better correlated with the simulated outputs than the inputs are with the simulated outputs. Given two mass maps, the normalized cross-correlation is given by
\begin{equation}
    r(k) = \frac{P_\mathrm{sim \times gen}(k)}{\sqrt{P_\mathrm{sim}(k){P_\mathrm{gen}(k)}}},
    \label{eq:ncc}
\end{equation}
where the numerator is the cross-correlation power spectrum of the two maps and the denominator is the geometric mean of their auto-correlations. The maps are perfectly correlated if $r=1$, uncorrelated if $r=0$, and perfectly anti-correlated if $r = -1$.

Figure~\ref{fig:cc} shows the pairwise cross correlation between the simulated output (ground truth) with the inputs and the generated output in blue. This curve gives us a metric on how well the reconstructed mass maps, across all scales, agree with the simulation mass maps. Plotted in the orange and purple curves are the correlation between the simulated SZ and X-ray inputs with the simulated outputs, respectively, which is a proxy for the performance of a linear bias model, which is effectively a baseline model for mapping the observables to the underlying densities. This graph serves to compare the spectral reconstruction performance of our model with the optimal performance of a linear bias model.

For the correlated simulation inputs and outputs, the model behaves similarly for both high-mass and low-mass clusters: the gas is better correlated with the input SZ and X-ray observables than the dark matter since the construction of the gas maps is much more closely aligned with the actual observables than the underlying dark matter maps, which have a much more complex relationship. More specifically, for the gas maps, SZ correlations are stronger than those of X-ray, which agree with Equations~\ref{eq:xry} and~\ref{eq:sz}: because SZ is proportional to pressure and density while X-ray is proportional to the square of density, we can expect the X-ray signal to be less correlated as the square will be much more sensitive to the small-scale clumpiness of the gas maps. Finally, the diffusion results in blue show that our model is able to learn a much stronger correlation across all scales. While we are able to achieve high correlations up to around $k = 4$ for dark matter, below this scale we lack sufficient constraining power on the substructure from the observables. To ameliorate this, the diffusion model learns to suggest plausible substructures which, while not exactly the true configuration observed in our simulations, are physically consistent with the larger structures. Thus, sometimes these substructures will be correct, and at other times they may be wrong, and the overall net effect will reduce the cross-correlations at smaller scales. \cite{Andres_2024} reconstructs the generated and simulated power spectrum and shows agreement by comparing the relative error, but does not compute $b(k)$ or $r(k)$, which we focus on to show the improvements of our model over linear models in the spectral domain.

In Table~\ref{tab:cc}, we summarize the average values of Figures~\ref{fig:bias} and~\ref{fig:cc}. Our bias values are close in agreement with unity across all scales for both gas and dark matter maps. Our cross-correlation values are close to unity at large scales, and around $0.97$ for gas and $0.80$ for dark matter for small scales, indicating that our model is able to reproduce the structure of clusters very well at these two different scale lengths.

\begin{table*}[htbp]
\centering
\begin{tabular}{ll|cccc} 
\toprule
\textbf{Cluster Size} & \textbf{Mass Map}  & \textbf{Small scale $b(k)$} & \textbf{Large scale $b(k)$} & \textbf{Small scale $r(k)$} & \textbf{Large scale $r(k)$} \\
\midrule
high mass & Gas           & $0.974_{-0.003}^{+0.004}$ & $0.988_{-0.007}^{+0.009}$  & $0.970_{-0.006}^{+0.008}$ & $0.994_{-0.005}^{+0.004}$ \\
          & Dark Matter   & $0.828_{-0.106}^{+0.096}$ & $0.987_{-0.010}^{+0.007}$  & $0.791_{-0.137}^{+0.125}$ & $0.982_{-0.015}^{+0.015}$\\
\midrule
low mass  & Gas           & $0.982_{-0.003}^{+0.005}$ & $0.989_{-0.005}^{+0.004}$  & $0.975_{-0.007}^{+0.008}$ & $0.995_{-0.004}^{+0.003}$ \\
          & Dark Matter   & $0.860_{-0.087}^{+0.077}$ & $0.972_{-0.008}^{+0.009}$ & $0.831_{-0.114}^{+0.101}$ & $0.986_{-0.011}^{+0.012}$ \\
\bottomrule
\end{tabular}
%\vspace{9pt}
\caption{Mean bias and cross-correlation values for the generated dark matter and gas maps per scale length bin, dependent on the galaxy cluster size (either high mass or low mass) and the scale length (either $k < 2$ h/Mpc for large scale or $k \geq 2$ h/Mpc for small scales). Across all of these classes, we achieve a close-to-unity bias and high correlation per scale bin using the diffusion model. This indicates that our model is learning a complex, nonlinear mapping between the SZ and X-ray inputs and the dark matter and gas outputs.} 
\label{tab:cc}
%\vspace{9pt}
\end{table*}
\section{Conclusion} \label{sec:conclusion}

We present a novel approach to map SZ and X-ray inputs of galaxy clusters to the gas and dark matter density maps using score-based conditional generative models. In particular, our model takes in SZ and X-ray images as conditional inputs, and combines that with a random sample from a Gaussian random field to produce a corresponding realization of the gas and dark matter maps. Our diffusion model uses Langevin dynamics sampling to generate these maps, wherein the model utilizes the learned score of the data distribution and evolves the noisy image through conditional reverse stochastic differential equations.

We trained our diffusion model using mock SZ and X-ray observations and simulated gas and dark matter densities from \texttt{HYPER}, reserving part of the dataset for downstream testing. During training, the model learns to match the score by reconstructing a gas and dark matter map from a random noisy image sample using the corresponding SZ and X-ray conditional input. As the model optimizes the score over time, it will better reconstruct the score and thus sample better from the posterior, generating more accurate realizations of the gas and dark matter maps.

Our diffusion model is able to generate highly realistic and accurate realizations of gas and dark matter maps. The samples from the model converge since a ten-sample mean is able to achieve over one e-folding of a two-sample variance for our test dataset. In addition, because the dataset is normalized by a global maximum and minimum so as to preserve relative intensities between clusters, the model is able to accurately predict the matter distribution across differently-sized clusters. Another indication that our model is able to distinguish between differently-sized clusters is that it is able to reconstruct the mean and spread of both the gas and dark matter density profiles to within 5\% on all spatial scales and all cluster sizes. Finally, in the spectral domain, the model achieves a close-to-unity agreement for the bias coefficients and high cross-correlation coefficients between the simulated and generated density maps, implying that our model learns a strong nonlinear mapping that can accurately probe both large and small scale cluster structures.

Future work can be done to improve the reconstruction of the mass maps in both the spatial and spectral domains. Currently, our loss is only the score-matching term, which does not include any information on the physical constraints of our data. We can utilize physics-informed losses during learning, where we augment our loss with extra terms, such as constraints on the total mass of the cluster or regularization terms on the bias and cross-correlation coefficients at different scale lengths. Improvements can also be made on the data end: by adding realistic and structured noise to our dataset, we can move towards using our score models to predict unknown density maps using observational X-ray and SZ images. The noising procedure of the score models will have to be subsequently adapted such that $\sigma_{min}$ matches the level of the inherent data noise, and consequently we expect our constraining power to be worse when we add noise, allowing us only to focus on high signal-to-noise systems and recovering larger-scale features.

Additional work can be done to analyze the dependency of these diffusion models on different hydrodynamical simulations. Since our mock observations of halos and baryon fractions are cosmology dependent, future work will to be done to test the diffusion model dependence on cosmological parameters. Finally, to shift our models towards real observations, we plan to train on data generated from simulations with AGN feedback such as IllustrisTNG and The Three Hundred Projects \citep[e.g.][]{Nelson_2019, Cui_2018}, as feedback may weaken the correlation between the densities and the observed SZ and X-ray measurements. We also plan to use weak lensing measurements as an additional conditional measurement: since it is largely AGN independent as it depends on total matter, most of which is dark matter, it will act as a feedback regularizer in the diffusion model training.  

The success of our diffusion model to accurately predict gas and dark matter realizations by sampling from a learned distribution has huge implications in real-data inference: these diffusion models can ultimately be used to predict unknown gas and dark matter distributions of real galaxy clusters of interest.

\begin{acknowledgements}
We thank Jun-Yan Zhu for discussions on generative modeling, and Nick Gnedin, Yuuki Omori, and Qirong Zhu for discussions on \texttt{HYPER}. We are grateful to Alexandre Adam for making his implementation of score models publicly available. We thank Jean-Baptiste Melin for providing cluster pressure profiles, and Erwin Lau and Daisuke Nagai for the tabulated cluster profiles from the TNG300 simulation. A.H. thanks Ashley Villar for guidance on parts of this project during the Spring 2024 ASTRON205 course at Harvard University. The $\texttt{HYPER}$ simulation was run on Endeavour at the NASA Advanced Supercomputing (NAS) Division. The machine learning work was run on the FASRC Cannon cluster supported by the FAS Division of Science Research Computing Group at Harvard University. This project is supported by NASA grant 80NSSC22K0821.
\end{acknowledgements}

\newpage

\bibliography{references}{}
\bibliographystyle{aasjournal}

\begin{appendix}

\section{Halo Profiles} \label{sec:haloprofiles}

\begin{figure}[h]
\includegraphics[width=\linewidth]{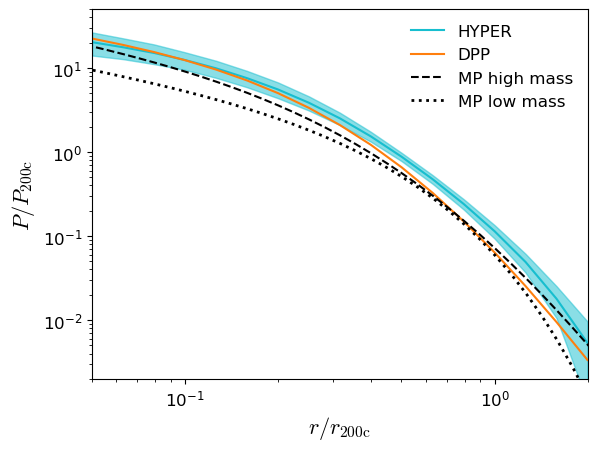}
\caption{The average (cyan) pressure profile and $1\sigma$ dispersion (shaded) for a subsample of galaxy clusters from the $\texttt{HYPER}$ simulation with $M_\mathrm{200m} \approx 6 \times 10^{14}\, M_\odot$ and $M_\mathrm{200c} \approx 4 \times 10^{14}\, M_\odot$. The simulated profiles agree with the DPP \citep[orange;][]{2021ApJ...908...91H} to within $\sim 20\%$. Additional, there is good agreement with the best-fit profiles for the low-mass (dotted) and high-mass (dashed) subsamples from \citet{2023A&A...678A.197M}.}
\label{fig:pprofile}
\end{figure}

The $\texttt{HYPER}$ simulation uses the debiased pressure profile \citep[DPP;][]{2021ApJ...908...91H} in the subgrid model for the ICM. To compare the simulation with the model, we use a subsample of halos positioned between our low-mass and high-mass ranges. Note that $\texttt{HYPER}$ outputs radial profiles based on the halo mass $M_\mathrm{200c}$ within radius $R_\mathrm{200c}$, where the average density is 200 times the \textit{critical} density. The pressure has been normalized by the characteristic pressure \citep[e.g.][]{2015ApJ...806...68L},
\begin{equation}
    P_\mathrm{200c} = \frac{GM_\mathrm{200c}}{2R_\mathrm{200c}}(200\rho_\mathrm{crit})f_\mathrm{b} ,
\end{equation}
where $f_\mathrm{b} = \Omega_\mathrm{b}/\Omega_\mathrm{m}$ is the cosmic baryon fraction. Figure \ref{fig:pprofile} shows that the simulated pressure profiles agree with the DPP to within $\sim 20\%$ over the radial range of interest. For comparison, we also show the best-fit profiles for the low-mass and high-mass subsamples from \citet{2023A&A...678A.197M}, based on joint measurement from Planck and SPT-SZ.

\begin{figure}[b]
\includegraphics[width=\linewidth]{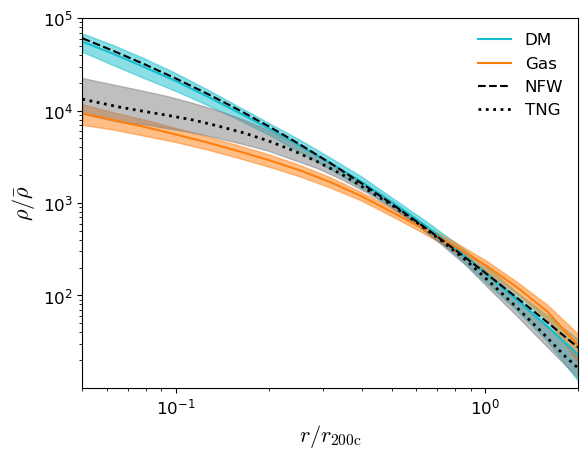}
\caption{The average (solid) density profile and $1\sigma$ dispersion (shaded) for a subsample of galaxy clusters from the $\texttt{HYPER}$ simulation with $M_\mathrm{200m} \approx 6 \times 10^{14}\, M_\odot$ and $M_\mathrm{200c} \approx 4 \times 10^{14}\, M_\odot$. The dark matter (cyan) density is in excellent agreement with the NFW profile (dashed). The gas density (orange) shows overall good agreement with the TNG300 simulation (gray) despite the different ICM modeling.}
\label{fig:dprofile}
\end{figure}

Figure \ref{fig:dprofile} shows the density profiles for the same subsample of simulated galaxy clusters. The dark matter density is in excellent agreement with the NFW universal density profile \citep{1997ApJ...490..493N}. The gas density is suppressed relative to the dark matter at small radii due to pressure support. Despite the different ICM modeling, there is also overall good agreement with the TNG300 simulation \citep[e.g.][]{2018MNRAS.481.1809B}. The \texttt{HYPER} gas densities are slightly lower at small radii, likely due to having higher central pressure.

\end{appendix}

\end{document}